\documentclass[11pt]{article}
\usepackage[hypertex]{hyperref}
\usepackage[dvipdfmx]{graphicx} 
\usepackage{amsmath,amssymb,amsfonts,bm,cite} 

\def\a{\alpha} \def\b{\beta}   \def\d{\delta} \def\D{\Delta}    \def\th{\theta}    \def\L{\Lambda} \def\m{\mu} \def\n{\nu}        \def\t{\tau} \def\ph{\phi}      \def\O{\Omega}

\def\dg{\dagger}  \def\nn{\nonumber}

\newcommand{\meV}{ {\rm meV} } \newcommand{\eV}{ {\rm eV} }  \newcommand{\MeV}{ {\rm MeV} } \newcommand{\GeV}{ {\rm GeV} } 

\def\ds{\displaystyle}

\newcommand{\lsp}{ \left ( } \newcommand{\rsp}{ \right ) } \newcommand{\Lg}{\mathcal{L}}     

\newcommand{\vev}[1]{ \langle {#1} \rangle }

  \newcommand{\Det}{{\rm Det}}

\newcommand{\Diag}[3]{ \begin{pmatrix} #1 & 0 & 0 \\ 0 & #2 & 0 \\ 0 & 0 & #3 \\\end{pmatrix}}

\usepackage{color}

\begin{document}

\begin{flushright}
STUPP-19-240
\end{flushright}

\vskip 1.35cm

\begin{center}
{\Large \bf Diagonal reflection symmetries and universal four-zero texture}

\vskip 1.2cm

Masaki J. S. Yang

\vskip 0.4cm

{\it Department of Physics, Saitama University, \\
Shimo-okubo, Sakura-ku, Saitama, 338-8570, Japan\\
}




\begin{abstract} 

In this paper,  
we consider a set of new symmetries in the SM: {\it diagonal reflection} symmetries $R \, m_{u,\nu}^{*} \, R = m_{u,\nu}, ~ m_{d,e}^{*} = m_{d,e}$ with $R =$ diag $(-1,1,1)$.
These generalized $CP$ symmetries predict 
the Majorana phases to be $\alpha_{2,3} /2 \sim 0$ or $\pi /2$. 
Realization of symmetries implies 
a broken chiral $U(1)_{\rm PQ}$ symmetry only for the first generation.
The axion scale is suggested to be 
$\langle {\theta_{u,d}} \rangle \sim \Lambda_{\rm GUT} \, \sqrt{m_{u,d} \, m_{c,s}} / v \sim 10^{12} \, $[GeV].
By combining the symmetries with the four-zero texture, 
the mass eigenvalues and mixing matrices of quarks and leptons are reproduced well. 
This scheme predicts the normal hierarchy, the Dirac phase $\d_{CP} \simeq 203^{\circ},$  and $|m_{1}|  \simeq 2.5$ or $6.2 \, $[meV].
In this scheme, the type-I seesaw mechanism and a given neutrino Yukawa matrix $Y_{\n}$ 
completely determine the structure of the right-handed neutrino mass $M_{R}$.  
A $u-\nu$ unification predicts the mass eigenvalues to be $ (M_{R1} \, , M_{R2} \, , M_{R3}) = (O (10^{5}) \, , O (10^{9}) \, , O (10^{14})) \, $[GeV].

\end{abstract} 

\end{center}

\section{Introduction}

The discovery of the neutrino oscillation \cite{Fukuda:1998mi, Ahmad:2001an}
proved the finite mass and mixing of neutrinos. 
To explain the peculiar mixing pattern, 
many flavor structures based on some symmetry such as 
four-zero texture 
\cite{Fritzsch:1995nx, Nishiura:1999yt, Matsuda:1999yx, Fritzsch:1999ee, Fritzsch:2002ga, Xing:2003zd, Xing:2003yj, Bando:2004hi, Matsuda:2006xa, Ahuja:2007vh, Xing:2015sva}, 
democratic texture \cite{Harari:1978yi, Koide:1983qe, Koide:1989zt, Tanimoto:1989qh, Fritzsch:1989qm, Lehmann:1995br, Fukugita:1998vn, Tanimoto:1999pj, Haba:2000rf, Hamaguchi:2002vi, Kakizaki:2003fc, Kobayashi:2004ha, Fritzsch:2004xc, Jora:2006dh, Mondragon:2007af, Xing:2010iu, Zhou:2011nu, Canales:2012dr, Yang:2016esx, Yang:2016crz}, 
$\m - \t$ symmetry 
\cite{Fukuyama:1997ky,Lam:2001fb,Ma:2001mr,Balaji:2001ex,Koide:2002cj,Kitabayashi:2002jd,Koide:2003rx,Ghosal:2003mq,Aizawa:2004qf,Ghosal:2004qb,Mohapatra:2005yu,Koide:2004gj,Kitabayashi:2005fc,Haba:2006hc,Xing:2006xa,Ahn:2006nu,Joshipura:2005vy,GomezIzquierdo:2009id,He:2011kn,He:2012yt,Gomez-Izquierdo:2017rxi, Fukuyama:2017qxb}, 
and $\m-\t$ reflection symmetry 
\cite{Harrison:2002et,Grimus:2003yn, Grimus:2005jk, Farzan:2006vj, Joshipura:2007sf, Adhikary:2009kz, Joshipura:2009tg, Xing:2010ez, Ge:2010js, Gupta:2011ct, Grimus:2012hu, Joshipura:2015dsa, Xing:2015fdg, He:2015afa, Chen:2015siy, He:2015xha, Samanta:2017kce, Xing:2017cwb, Nishi:2018vlz, Nath:2018hjx, Sinha:2018xof, Xing:2019edp, Pan:2019qcc}, 
have been studied. 
However, these symmetries often have large corrections of symmetry breaking on the order of $\sim O(0.1)$. 
Among them, $\m-\t$ reflection symmetries for quarks and leptons have been recently discussed \cite{Yang:2020qsa}.


In this paper, we consider a set of new symmetries with the accuracy of $\simeq O(2,3 \%)$ 
in the Standard Model (SM), i.e., diagonal reflection symmetries for quarks and leptons. 
The previous study of $\m-\t$ reflection symmetries are
 translated to forms $R \, m_{u,\n}^{*} \, R = m_{u,\n}, ~ m_{d,e}^{*} = m_{d,e}$ with $R =$ diag $(-1,1,1)$ 
by a redefinition of fermion fields. 
We call such a symmetry {\it diagonal reflection} 
because it is a diagonal remnant of $\mu-\tau$ reflection symmetry after deduction of $\mu-\tau$ symmetry.  
Each of them is just a generalized $CP$ (GCP) symmetry 
\cite{Ecker:1980at,Ecker:1983hz, Gronau:1985sp, Ecker:1987qp,Neufeld:1987wa,Ferreira:2009wh,Feruglio:2012cw,Holthausen:2012dk,Ding:2013bpa,Girardi:2013sza,Nishi:2013jqa,Ding:2013hpa,Feruglio:2013hia,Ding:2014ora,Ding:2014hva,Chen:2014tpa,Li:2015jxa,Turner:2015uta,Penedo:2017vtf,Nath:2018fvw}
and no longer a $\m-\t$ reflection. 

The form of the symmetries suggests that the flavored $CP$ violation only comes from 
a chiral symmetry breaking of the first generation. 
As a justification of  
diagonal reflection symmetries and a zero texture $(m_{f})_{11} = 0$,  
simultaneous breaking of a chiral $U(1)_{\rm PQ}$ \cite{Peccei:1977hh} and a generalized $CP$ symmetry is discussed 
in a specific two Higgs doublet model (2HDM).
As a result, an invisible (flavored) axion \cite{Kim:1979if,Shifman:1979if, Zhitnitsky:1980tq,Dine:1981rt, Davidson:1981zd, Wilczek:1982rv, Berezhiani:1989fp, Ahn:2014gva}
 (a {\it flaxion} \cite{Ema:2016ops} or {\it axiflavon} \cite{Calibbi:2016hwq}) 
appears in conjunction with solving the strong CP problem  \cite{tHooft:1976fv}.
The axion scale is suggested to be $\vev{\th_{u,d}} \sim \L_{\rm GUT} \, \sqrt{m_{u,d} \, m_{c,s}} / v \sim 
10^{12} \, $[GeV].  This value can produce  the dark matter abundance $\O_{a} h^{2} \sim 0.2$ and is very intriguing. 
It is also applicable to a solution of the strong CP problem using the discrete symmetry $P$ \cite{Mohapatra:1978fy,Beg:1978mt} or $CP$ \cite{Nelson:1983zb} because the diagonal reflection symmetries can reconcile the CKM phase $\d_{\rm CKM}$ and $\th_{\rm QFD}^{\rm tree} = {\rm Arg} \, \Det[m_{u} m_{d}] = 0$ without Hermiticity or mirror fermions \cite{Barr:1991qx}.

An additional assumption $(m_{\n})_{13} = 0$ (which can be justified by Eq.~(\ref{13zero}) in the left-right symmetric models \cite{Pati:1974yy,Senjanovic:1975rk,Mohapatra:1974hk}) realizes 
diagonal reflection with universal four-zero texture, which restricts fermion mass matrices to have only four parameters. 
This scheme provides proper masses, mixing, and CP phases of quarks and leptons. 
It predicts the Dirac phase $\d_{CP} \simeq 203^{\circ}$, 
the Majorana phases $(\a_{2}, \a_{3}) \simeq (11.3^{\circ}, 7.54^{\circ})$ up to $180^{\circ}$, 
the normal mass hierarchy, and the lightest neutrino mass $|m_{1}| \simeq 2.5$ or $6.2 \, [\meV]$.


The main purpose of this paper is to constrain the mass matrix of right-handed neutrinos $M_{R}$ 
using the diagonal reflection symmetries, the four-zero texture, and the type-I seesaw mechanism \cite{Minkowski:1977sc, Yanagida:1979as, Mohapatra:1979ia, GellMann:1980v}. 
The matrix $M_{R}$ also exhibits diagonal reflection symmetry  with a four-zero texture 
because four-zero textures are type-I seesaw invariant \cite{Nishiura:1999yt,Fritzsch:1999ee}. 
For a given neutrino Yukawa matrix $Y_{\n}$, 
the texture of $M_{R}$ is completely determined by the 
seesaw mechanism  in this scheme. 
A $u-\nu$ unification predicts the mass eigenvalues to be 
$ (M_{R1} \, , M_{R2} \, , M_{R3}) = (O (10^{5}) \, , O (10^{9}) \, , O (10^{14})) \, $[GeV].

Quantum corrections hardly break these symmetries 
because couplings of the first generation are very small. 
A qualitative analysis shows that the symmetries are retained as approximate ones under the renormalization group equations of the SM. 


This paper is organized as follows. 
The next section gives the definition of diagonal reflection symmetries.
Sec.~3 discusses a realization of diagonal reflection symmetries and implications regarding the strong $CP$ problem. 
Sec.~4 presents an analysis of physical parameters and universal four-zero texture.
In Sec.~5, we discuss stability under quantum corrections.
The final section is devoted to a summary.

\section{Diagonal reflection symmetries} 
To start, we show a new set of symmetries.
The mass matrices of the SM fermions $f= u,d,e,$ and neutrinos $\n_{L}$ are defined by 
\begin{align}
\Lg \ni  \sum_{f} -  \bar f_{Li } m_{f ij}^{BM} f_{Rj} - \bar \n_{L i} m_{\n ij}^{BM} \n_{L j}^{c} + {\rm h.c.} \, .
\end{align}
Here, we assume Hermitian $m_{f}^{BM}$ and complex-symmetric $m_{\n}^{BM}$, 
which can produce successful mass eigenvalues and mixing matrices $V_{\rm CKM}$ and $U_{\rm MNS}$ \cite{Yang:2020qsa};
\begin{align}
m_{u}^{BM} & = 
\begin{pmatrix}
 0 & -\frac{{C_{u}}}{\sqrt{2}} & -\frac{{C_{u}}}{\sqrt{2}} \\
- \frac{{C_{u}}}{\sqrt{2}} & \frac{{\tilde B_{u}}}{2}+\frac{{A_{u}}}{2} & \frac{{\tilde B_{u}}}{2}-\frac{{A_{u}}}{2}-i{B_{u}} \\
- \frac{{C_{u}}}{\sqrt{2}} &\frac{{\tilde B_{u}}}{2}-\frac{{A_{u}}}{2} +  i {B_{u}} & \frac{{\tilde B_{u}}}{2}+\frac{{A_{u}}}{2} \\
\end{pmatrix} ,  
\label{mup}
\\
m_{d}^{BM} & = 
\begin{pmatrix}
 0 & \frac{i {C_{d}}}{\sqrt{2}} & \frac{i {C_{d}}}{\sqrt{2}} \\
 -\frac{i {C_{d}}}{\sqrt{2}} & \frac{\tilde B_{d}}{2}+\frac{{A_{d}}}{2} &\frac{\tilde B_{d}}{2}-\frac{{A_{d}}}{2}  -i {B_{d}}\\
 -\frac{i {C_{d}}}{\sqrt{2}} & \frac{\tilde B_{d}}{2}-\frac{{A_{d}}}{2} + i  {B_{d}} & \frac{\tilde {B_{d}}}{2}+\frac{{A_{d}}}{2} \\
\end{pmatrix} ,
 \label{mdown}
\end{align}
and 
\begin{align}
m_{\n}^{BM} &
= 
\begin{pmatrix}
 -a_{\n} & {1\over \sqrt{2}} (b_{\n} - i c_{\n}) &  {1\over \sqrt{2}} (b_{\n} + i c_{\n})  \\
 {1\over \sqrt{2}} (b_{\n} - i c_{\n})  & \frac{f_{\n}}{2} -\frac{d_{\n}}{2}+ i e_{\n} & -\frac{f_{\n}}{2}-\frac{d_{\n}}{2} \\
 {1\over \sqrt{2}} (b_{\n} + i c_{\n})  & -\frac{f_{\n}}{2}-\frac{d_{\n}}{2} & \frac{f_{\n}}{2} -\frac{d_{\n}}{2}- i e_{\n}\\
\end{pmatrix} \label{mnu}
, \\
m_{e}^{BM} &=
\begin{pmatrix}
 0 & \frac{i {C_{e}}}{\sqrt{2}} & \frac{i {C_{e}}}{\sqrt{2}} \\
 -\frac{i {C_{e}}}{\sqrt{2}} & \frac{\tilde B_{e}}{2}+\frac{A_{e}}{2} &\frac{\tilde B_{e}}{2}-\frac{A_{e}}{2}  -i {B_{e}}\\
 -\frac{i {C_{e}}}{\sqrt{2}} & \frac{\tilde B_{e}}{2}-\frac{A_{e}}{2} + i  {B_{e}} & \frac{\tilde B_{e}}{2}+\frac{A_{e}}{2} \\
\end{pmatrix} . \label{mbe}
\end{align}
The hermiticity of Yukawa matrices is justified by the parity symmetry in the left-right symmetric models 
\cite{Pati:1974yy,Senjanovic:1975rk,Mohapatra:1974hk}.
These matrices~(\ref{mup})-(\ref{mbe}) separately satisfy $\m - \t$ reflection symmetries \cite{Harrison:2002et,Grimus:2003yn}:
\begin{align}
T_{u} (m_{u,\n}^{BM})^{*} T_{u} = m_{u,\n}^{BM}  , ~~~ 
T_{d} (m_{d,e}^{BM})^{*} T_{d} = m_{d,e}^{BM}  , ~~~ 
\label{mutausym}
\end{align} 
where 
\begin{align}
T_{u}= 
\begin{pmatrix}
 1 & 0 & 0 \\
 0 & 0 & 1 \\
 0 & 1 & 0 \\
\end{pmatrix} , ~~~
T_{d} =
\begin{pmatrix}
 1 & 0 & 0 \\
 0 & 0 & -1 \\
 0 & -1 & 0 \\
\end{pmatrix} .
\end{align}
In general, a Hermitian or complex-symmetric matrix with a $\m-\t$ reflection symmetry 
has six parameters. 
Eq.~(\ref{mnu}) is a general complex-symmetric matrix which satisfies Eq.~(\ref{mutausym}). 
Eq.~(\ref{mup}), Eq.~(\ref{mdown}), and Eq.~(\ref{mbe}) have four parameters 
with two additional constraints, $(m_{f})_{11} = 0$ and $(m_{f})_{12} = (m_{f})_{13}$.

A simultaneous redefinition of all fermion fields $f' = U_{BM} f$ and $\n' = U_{BM} \n$ by
the following bi-maximal transformation $U_{BM}$, 
\begin{align}
m_{f} &\equiv  U_{BM} m_{f}^{BM} U_{BM}^{\dag}, ~~
m_{\n} \equiv U_{BM} m_{\n}^{BM} U_{BM}^{T} , 
~~~U_{BM} \equiv
\begin{pmatrix}
 1 & 0 & 0 \\
 0 & \frac{i}{\sqrt{2}} & \frac{i}{\sqrt{2}} \\
 0 & -\frac{1}{\sqrt{2}} & \frac{1}{\sqrt{2}} \\
\end{pmatrix} , 
\label{bimaximal}
\end{align}
leads to Hermitian four-zero textures \cite{Fritzsch:1995nx}
 and a symmetric neutrino mass; 
\begin{align}
m_{u} &= 
\Diag{i}{1}{1}
\begin{pmatrix}
0 & C_{u} & 0 \\ 
 C_{u} &\tilde B_{u} & B_{u} \\
0 & B_{u} & A_{u} 
\end{pmatrix} 
\Diag{-i}{1}{1} ,  \label{mu} ~~~
m_{d}  =  
\begin{pmatrix}
0 & C_{d} & 0 \\ 
C_{d} &\tilde B_{d} & B_{d} \\
0 & B_{d}  & A_{d} 
\end{pmatrix} ,
 \\
m_{\n} & = 
\Diag{-i}{1}{1}
\begin{pmatrix}
a_{\n} & b_{\n} & c_{\n} \\ 
b_{\n} & d_{\n} & e_{\n} \\
c_{\n} & e_{\n}  & f_{\n} 
\end{pmatrix} \label{mn}
\Diag{-i}{1}{1}
, ~~~
m_{e} =
\begin{pmatrix}
0 & C_{e} & 0 \\ 
C_{e} &\tilde B_{e} & B_{e} \\
0 & B_{e}  & A_{e} 
\end{pmatrix} . 
\end{align}
Here,  $a_{\n}\sim f_{\n}$ and $A_{f} \sim C_{f}$ are real parameters 
that satisfy $A_{f} > \tilde B_{f} > B_{f} \gg C_{f}$. 
In this basis, the assumptions are deformed to be $(Y_{f})_{11}, (Y_{f})_{13}, (Y_{f})_{31} = 0$ for $f = u,d,e$. 
%
We will partially discuss a justification of the texture later. 
Note that a $\m-\t$ reflection symmetry is not imposed on $m_{\n}$~(\ref{mn}). 

In this basis of the four-zero texture,  
the $\m-\t$ reflection symmetries~(\ref{mutausym}) 
are rewritten as
\begin{align}
 U_{BM} T_{u,d} U_{BM}^{T} m_{u,d}^{*} U_{BM}^{*} T_{u,d} U_{BM}^{\dg} =  m_{u,d}. 
\end{align}
Surprisingly, 
\begin{align}
- & U_{BM}^{*} T_{u} U_{BM}^{\dg} = \Diag{-1}{1}{1} \equiv R , \\
& U_{BM}^{*} T_{d} U_{BM}^{\dg} = \Diag{1}{1}{1} = 1_{3} . 
\end{align}
Then, the $\m-\t$ reflection symmetries in the four-zero basis are transformed into 
\begin{align}
R m_{u,\n}^{*} R = m_{u,\n} , ~~~ 
m_{d,e}^{*} = m_{d,e}. 
\label{refsym}
\end{align}
Hermitian or symmetric mass matrices that satisfy Eq.~(\ref{refsym}) are given by 
\begin{align}
m_{u} &= 
\begin{pmatrix}
a_{u} & i b_{u} & i c_{u} \\
- i b_{u} & d_{u} & e_{u} \\
- i c_{u} & e_{u} & f_{u}
\end{pmatrix} , 
~
m_{\n} = 
\begin{pmatrix}
a_{\n} & i b_{\n} & i c_{\n} \\
 i b_{\n} & d_{\n} & e_{\n} \\
 i c_{\n} & e_{\n} & f_{\n}
\end{pmatrix} , 
~ 
m_{d,e} = 
\begin{pmatrix}
a_{d,e} & b_{d,e} & c_{d,e} \\
b_{d,e} & d_{d,e} & e_{d,e} \\
c_{d,e} & e_{d,e} & f_{d,e}
\end{pmatrix} ,
\end{align}
with real parameters $a_{f} \sim f_{f}$. 
The mass matrices~(\ref{mu})-(\ref{mn}) 
certainly satisfy these conditions. 
We call such a symmetry {\it diagonal reflection} 
because it is a diagonal remnant of $\mu-\tau$ reflection symmetry after deduction of $\mu-\tau$ symmetry.  
Each of them is just a generalized $CP$ symmetry \cite{Ecker:1983hz,Ecker:1987qp,Neufeld:1987wa,Ferreira:2009wh,Holthausen:2012dk} and no longer a $\m-\t$ reflection. 
The textures~(\ref{mu}) are discussed for quarks and CKM matrices in many studies (\cite{Xing:2003yj} and references therein). 
However, we cannot find a report that indicates the existence of GCP symmetries.

The latest calculation shows an example of Yukawa matrices compatible with all the flavor data of quarks  \cite{Xing:2015sva}:
\begin{align}
Y_{u}^{0} & \simeq {0.9 m_{t} \sqrt 2 \over v}
\begin{pmatrix}
0 & 0.0002 \, i & 0 \\ 
-0.0002 \, i & 0.10 & 0.31 \, e^{ \pm 0.02 \pi} \\
0 & 0.31 \, e^{ \mp 0.02 \pi} & 1
\end{pmatrix} 
 ,  \label{bestfit1} \\
Y_{d}^{0} & \simeq {0.9 m_{b} \sqrt 2 \over v}
\begin{pmatrix}
0 & 0.005 & 0 \\ 
0.005 & 0.13 & 0.31 \, e^{ \mp 0.02 \pi} \\
0 & 0.31 \, e^{ \pm 0.02 \pi}  & 1
\end{pmatrix} 
 , 
\label{bestfit2}
\end{align}
where $v = 246 \, [\GeV]$ is the vacuum expectation value (vev) of the SM Higgs field. 
The textures~(\ref{mu}) agree with (\ref{bestfit1}) and (\ref{bestfit2}) with an accuracy of $O(2,3 \%)$. 
Breaking effects come from the phases of the 23 element $B_{u,d} \, e^{i \varphi_{u,d}}$, where 
$\varphi_{u,d} \sim  \pm 0.02 \pi$. 

Because the conditions~(\ref{refsym})
depend on a basis, they are changed by further redefinitions of fermion fields
(the weak basis transformations \cite{Branco:1999nb,Branco:2007nn}). 
For example, rephasing of quark fields $Q = q,u,d$
\begin{align}
Q' = P_{Q}^{\dg} Q, ~~~  P_{Q}= {\rm diag} (e^{i \ph_{Q}}, 1 ,1),
\end{align}
leads to $CP$-violating quark masses $\tilde m_{u, d}$; 
\begin{align}
\tilde m_{u} &= P^{\dg}_{q} m_{u} P_{u} = 
\begin{pmatrix}
a_{u} & i e^{- i\ph_{q}} b_{u} & i e^{- i\ph_{q}} c_{u} \\
- i e^{i\ph_{u}} b_{u} & d_{u} & e_{u} \\
- i e^{i\ph_{u}} c_{u} & e_{u} & f_{u}
\end{pmatrix} , \label{tildemu}  \\
\tilde m_{d} &= P^{\dg}_{q} m_{d} P_{d} = 
\begin{pmatrix}
a_{d} & e^{- i\ph_{q}} b_{d} & e^{- i\ph_{q}} c_{d} \\
e^{i\ph_{d}} b_{d} & d_{d} & e_{d} \\
e^{i\ph_{d}} c_{d} & e_{d} & f_{d}
\end{pmatrix} . \label{tildemd}
\end{align}
In this case, using the following equivalent transformation 
\begin{align}
R_{q,u} \equiv P_{q,u} R P_{q,u} = \Diag{- e^{2 i \ph_{q,u}}}{1}{1} ,  ~~~ 
\tilde R_{q,d} \equiv P_{q,d} 1_{3} P_{q,d} = \Diag{+ e^{2 i \ph_{q,d}}}{1}{1} , 
\end{align}
deforms the diagonal reflection symmetries~(\ref{refsym}) as
\begin{align}
R_{q}^{\dg} \tilde m_{u}^{*} R_{u} = \tilde m_{u} , ~~~ 
\tilde R_{q}^{\dg} \tilde m_{d}^{*} \tilde R_{d} = \tilde m_{d} . \label{refsym2}
\end{align}
In this basis, the Hermiticity of the quark masses is lost, 
as shown in Eqs.~(\ref{tildemu}) and~(\ref{tildemd}). 
The symmetries in Eq.~(\ref{mutausym}), Eq.~(\ref{refsym}), and Eq.~(\ref{refsym2}) are all equivalent under redefinitions of fermion fields. 

\section{Realization of the symmetries}

The $\m - \t$ reflection symmetry is often realized as a remnant of a larger flavor symmetry, such as $A_{4},\, Z_{2} \times Z_{2}, \, U(1)_{L_{\m} - L_{\t}}$ 
\cite{Harrison:2002et,Grimus:2003yn, Grimus:2005jk, Farzan:2006vj, Joshipura:2007sf, Adhikary:2009kz, Joshipura:2009tg, Xing:2010ez, Ge:2010js, Gupta:2011ct, Grimus:2012hu, Joshipura:2015dsa, Xing:2015fdg, He:2015afa, Chen:2015siy, He:2015xha, Samanta:2017kce, Xing:2017cwb, Nishi:2018vlz, Nath:2018hjx, Sinha:2018xof, Xing:2019edp, Pan:2019qcc}. 
The origin of four-zero texture is also discussed in the $S_{3L} \times S_{3R}$ model \cite{Xing:1996hi, Kang:1997uv,Mondragon:1998gy,Barranco:2010we}. 
Thus, in this section, we concentrate on a realization of the diagonal reflection symmetries. 
Because Eq.~(\ref{mutausym}) or Eq.~(\ref{refsym}) imposes two independent GCP symmetries, 
the underlying CP should be broken separately in the up- and down-sector \cite{Ding:2013bpa}.

To this end, the following $U(1)_{\rm PQ} \times Z_{2}$ flavor symmetry and 
 a GCP symmetry are imposed on the 2HDM. 
 A similar model-building and its UV completion can be found in \cite{Shin:1985cg,Shin:1985vi,Kang:1985ua}. 
\begin{itemize}
\item $Z_{2}^{\rm NFC}$ : It realizes the natural flavor conservation (NFC) \cite{Glashow:1976nt} and prohibits 
 flavor changing neutral currents (FCNCs) by two Higgs doublets. 

\item $U(1)_{\rm PQ}$ : A chiral (PQ) symmetry \cite{Peccei:1977hh} that prohibits the mass of 
the first generation\footnote{ A discrete symmetry larger than $Z_{3}$ is also a possible choice.}. 
It is a kind of flavored PQ symmetry \cite{Davidson:1981zd, Wilczek:1982rv, Berezhiani:1989fp, Ahn:2014gva}.

\item $CP$ : A generalized $CP$ symmetry that restricts phases of Yukawa couplings. 
As an alternative,  
the driving field method \cite{Altarelli:2005yx}  is utilized to generate the relative phases. 
\end{itemize}
Two SM singlet flavon fields $\th_{u, d}$ are introduced to the 2HDM. 
These flavons have nontrivial charges under the $U(1)_{\rm PQ}$ and $CP$ symmetries.
Simultaneous breaking of these symmetries by vevs of $\th_{u,d}$ 
provokes CPV only for the first generation.
The charge assignment of fields is presented in Table 1. 

Under the $U(1)_{\rm PQ}$ symmetry, 
only the first-generation has nontrivial charges as
\begin{align}
q_{1 L}  \to e^{-i \a} q_{1L}, ~~ 
u_{1 R} &\to e^{i \a} u_{1R} , ~~ 
d_{1 R} \to e^{i \a} d_{1R} , ~~   \\
l_{1 L} \to e^{-i \a} l_{1L}, ~~ 
\n_{1 R} &\to e^{i \a} \n_{1R}, ~~ 
e_{1 R} \to e^{i \a} e_{1R} . 
\end{align}
The bilinear terms $\bar q_{Li} u_{Rj}, \bar q_{Li} d_{Rj}, \bar l_{Li} \n_{Rj}$ and, $\bar l_{Li} e_{Rj}$ (associated with Yukawa interactions) are transformed under $U(1)_{\rm PQ}$ as
\begin{align}
\lsp
\begin{array}{c|cc}
e^{ 2 i \a} & e^{i \a} & e^{i \a} \\ \hline
e^{i \a} & 1 & 1 \\
 e^{i \a} & 1 & 1 \\
\end{array}
\rsp .
\end{align}
\begin{table}[htb]
  \begin{center}
    \begin{tabular}{|c|ccccc|} \hline
           & $SU(2)_{L}$ & $U(1)_{Y}$ &  $Z_{2}^{\rm NFC}$ & $U(1)_{\rm PQ}$ & $CP$ \\ \hline \hline
      $q_{Li}$ & \bf 2 & $1/6$ & 1 & $-1,0,0$ & 1 \\
      $u_{Ri}$ & \bf 1 & $2/3$ & 1  & $1,0,0$ & 1\\ 
      $d_{Ri}$ & \bf 1 & $-1/3$ & $-1$ & $1,0,0$ & 1\\ 
      $l_{Li}$ & \bf 2 & $-1/2$ & 1 & $-1,0,0$ & 1\\
      $\n_{Ri}$ & \bf 1 & $0$ & 1 & $1,0,0$ & 1\\ 
      $e_{Ri}$ & \bf 1 & $-1$ &$-1$ & $1,0,0$ &  1 \\ \hline
      $H_{u}$ & \bf 2 & $-1/2$ & 1 & 0 & 1\\ 
      $H_{d}$ & \bf 2 & $1/2$ & $-1$ & 0  & 1 \\       
      $\th_{u}$ & \bf 1 & $1$ & 1 & $-1$ & $+i$ \\  
      $\th_{d}$ & \bf 1 & $1$ & $-1$ & $-1$ & $-i$\\ \hline
    \end{tabular}
    \caption{Charge assignments of the SM fermions and scalar fields under gauge and flavor symmetries.}
  \end{center}
\end{table}

Under these discrete symmetries, 
the most general Yukawa interactions are written as
\begin{align}
- \Lg &\ni  \bar q_{L} (\tilde Y_{u}^{0} + {\th_{u} \over \L} \tilde Y_{u}^{1} + {\th_{u}^{2} \over \L^{2}} \tilde Y_{u}^{2} + {\th_{d}^{2} \over \L^{2}} \tilde Y_{u}'{}^{2}  ) u_{R}  H_{u} 
\\ & +  \bar q_{L} (\tilde Y_{d}^{0} + {\th_{d} \over \L} \tilde Y_{d}^{1} + {\th_{u} \th_{d} \over \L^{2}} \tilde Y_{d}^{2} )d_{R} H_{d} + h.c. \, , 
\end{align}
where $\L$ is a cut-off scale. An analogous formula holds in the lepton sector.
The Yukawa matrices are parameterized as
\begin{align}
\tilde Y_{u,d}^{0} =
\begin{pmatrix}
0 & 0 & 0 \\ 
0 & \tilde d_{u,d} & \tilde c_{u,d} \\
0 & \tilde b_{u,d} & \tilde a_{u,d}
\end{pmatrix} ,  ~~~ 
\tilde Y_{u,d}^{1}  =
\begin{pmatrix}
0 & \tilde e_{u,d} & \tilde f_{u,d} \\ 
\tilde g_{u,d} & 0 & 0 \\
\tilde h_{u,d} & 0 & 0
\end{pmatrix} ,  ~~~ 
\end{align}
and $\tilde Y^{2}_{f}$ have only an 11 matrix element, 
which has a small influence. 
These Yukawa matrices satisfy the condition 
\begin{align}
(\tilde Y^{0}_{u,d})_{ij} \, (\tilde Y^{1}_{u,d})_{ij} = 0 ~~ ({\rm no~sum}), 
\label{nosum}
\end{align}
similar to consistency conditions of 
general parity (or $CP$) and flavor symmetry \cite{Ecker:1980at,Ecker:1983hz}. 

The generalized $CP$ invariance 
\begin{align}
\th_{u}^{*} = +i \th_{u}, ~~ \th_{d}^{*} = - i \th_{d} , ~~ \phi^{*} = \phi ~~ \text{ for other fields}
\end{align}
restricts relative complex phases of the matrix elements as
\begin{align}
(\tilde Y_{u,d}^{0})^{*} = \tilde Y_{u,d}^{0} ,  ~~~
\tilde Y_{u}^{1} = e^{i \pi/4} |\tilde Y_{u}^{1}| ,  ~~~ \tilde Y_{d}^{1} = e^{- i \pi/4} |\tilde Y_{d}^{1}| .
\end{align}

Next, we investigate the transformation properties of the Higgs potential.  
The potential can be written as 
\begin{align}
V = V^{1}(H_{u}, H_{d}) + V^{2}(H_{u,d}, \th_{u,d}) + V^{3}(\th_{u}, \th_{d}).
\end{align}
$V^{1}$ is obviously real because the GCP is the canonical $CP$ for the Higgs doublets $H_{u,d}$. 
Among bi-linear terms comprising $\th_{u}$ and $\th_{d}$, only $\th_{u}^{*} \th_{u}$ and $\th_{d}^{*} \th_{d}$ are
 invariant under $U(1)_{\rm PQ} \times Z_{2}^{\rm NFC}$
 (both $\th_{u}^{*} \th_{d}$ and its complex conjugate $\th_{d}^{*} \th_{u}$ have charge $-1$ under $Z_{2}^{\rm NFC}$ and $-1$ under $CP$). 
Then, $V_{2}$ has only real terms because $\th_{u}^{*} \th_{u}$ and $\th_{d}^{*} \th_{d}$ have trivial $CP$ charges.
 Finally, quartic terms made from the flavons should be a combination 
 between $\{ |\th_{u}|^{2} , |\th_{d}^{2}| \}$ or $\{ \th_{u}^{*} \th_{d} , \th_{d}^{*} \th_{u}\}$, 
 such as $|\th_{u}|^{2} |\th_{d}^{2}|$ or $\th_{u}^{*} \th_{d} \th_{u}^{*} \th_{d}$. 
Because these terms have trivial charges under $CP$, $V_{3}$ is GCP invariant, 
so the whole Higgs potential $V$ is invariant under $CP$. 
 Therefore, in this basis, CP phases are localized only in the first generation of Yukawa matrices. 
Real vevs of the flavon fields $\vev{\th_{u,d}}$ 
provokes a spontaneous symmetry breaking (SSB) of $U(1)_{\rm PQ}, Z_{2}^{\rm NFC},$ and $CP$.

As a result, 
the vevs $\vev{\th_{u,d}}$  produce the following textures 
\begin{align}
Y_{u,d} = (\tilde Y_{u,d}^{0} + {\vev{\th_{u,d}}\over \L} \tilde Y_{u,d}^{1}
 + {\vev{\th_{u,d}}^{2} \over \L^{2}} \tilde Y_{u,d}^{2}) 
= 
\begin{pmatrix}
O({\vev{\th_{u,d}}^{2} \over \L^{2}} )  &\tilde e \, {\vev{\th_{u,d}}\over \L} e^{i \varphi_{u,d}} &\tilde f {\vev{\th_{u,d}}\over \L} \, e^{i \varphi_{u,d}} \\[3pt]
\tilde g \, {\vev{\th_{u,d}}\over \L} e^{i \varphi_{u,d}} & \tilde d_{u,d} & \tilde c_{u,d} \\[3pt]
\tilde h \,{\vev{\th_{u,d}}\over \L} e^{i \varphi_{u,d}} & \tilde b_{u,d} & \tilde a_{u,d}
\end{pmatrix} ,
\label{fullYukawa}
\end{align}
where
\begin{align}
\varphi_{u} = + \pi/4 , ~~~ \varphi_{d} = - \pi/4 .
\label{phases}
\end{align}
These vevs can be estimated from the
best fit values for $Y_{u,d}$~(\ref{bestfit1}) and (\ref{bestfit2}) as 
\begin{align}
 {\vev{\th_{u}} \over \L} | \tilde Y_{u}^{1} | &\simeq {\sqrt{2 m_{u} \, m_{c}} \over v \, \sin \b}  \simeq {3 \times 10^{-4} \over \sin \b} ,  \label{flavonvev1} \\
  {\vev{\th_{d}} \over \L} | \tilde Y_{d}^{1} | &\simeq {\sqrt{2 m_{d} \, m_{s}} \over v \, \cos \b}  \simeq {1 \times 10^{-4} \over \cos \b} ,
 \label{flavonvev2}
 \end{align}
where $\vev{ H_{u}^{0}} \equiv v \sin \b / \sqrt 2,  \vev{H_{d}^{0}} \equiv v \cos \b / \sqrt 2 $ with $\vev{H_{u}^{0}}^{2} + \vev{ H_{d}^{0}}^{2} = v^{2}/2$.  
The small 11 matrix elements in Eq.~(\ref{fullYukawa}) are generated from $\tilde Y_{f}^{2}$. 
In many cases, 
they are negligible compared with the Yukawa eigenvalues of the first generation: 
\begin{align}
 {\vev{\th_{u,d}}^{2} \over \L^{2}}  \simeq {10^{-8} (\times \tan^{2} \b) \over | \tilde Y_{u,d}^{1} |^{2} }
~ \lesssim ~( y_{u} , y_{d}) \simeq ({m_{u} \over v \sin \b}, {m_{d} \over v \cos \b}) \simeq (10^{-5}, 10^{-5} \tan \b) . 
\end{align}
Therefore, Eq.~(\ref{fullYukawa}) and (\ref{phases}) satisfy the diagonal reflection symmetries~(\ref{refsym2}) 
with $\ph_{u} = 3\pi/4, ~ \ph_{q} = - \ph_{d} = \pi/4$, and $(m_{f})_{11} \simeq 0$. 

In this construction, 
Eqs.~(\ref{bestfit1}) and (\ref{bestfit2}) stand for 
$\tilde Y_{u}^{0} \simeq \tilde Y_{d}^{0}$ and $\tilde Y_{u}^{1} \sim \tilde Y_{d}^{1}$. 
This indicates the existence of $u-d$ unification, such as the left-right symmetric model. 
Moreover, with a $u-d$ unified relation $\tilde Y_{u}^{1} = \tilde Y_{d}^{1}$ (in the other basis of $CP$ phases),  
simultaneous rotation of 2-3 generations by a real orthogonal matrix $O_{23}$ can realize 
zero textures 
\begin{align}
(Y_{u})_{13} = (Y_{d})_{13} = (Y_{u})_{31} = (Y_{d})_{31} = 0.
\label{13zero}
\end{align}
Then, the four-zero textures with the diagonal reflection symmetries appear. 
Note that $O_{23}$ is commutative with 
the diagonal reflection symmetries because it satisfies $R \, O_{23}^{*} \, R = O_{23}$.

Realization of four-zero texture in the left-right symmetric model, 
such as a model in \cite{Xing:2015sva}, 
seems to lead to a more concise model. 
We leave this for future work. 

\subsection{Implications for the strong CP problem}

As a related issue, the strong $CP$ problem is considered \cite{tHooft:1976fv}.
This is a fine-tuning problem of $\bar \th  = \th_{\rm QCD} + \th_{\rm QFD}$, 
a sum of the QCD $\th$-term $\th_{\rm QCD}$ and its 
fermionic contribution $\th_{\rm QFD} = {\rm Arg} \, \Det[m_{u} m_{d}]$ \cite{Cheng:1987gp}. 

Although $Y_{u,d}$ in Eq.~(\ref{fullYukawa}) are not Hermitian matrices, 
$\th_{\rm QFD}^{\rm tree} = 0$ holds 
because they satisfy 
\begin{align}
\ph_{u}+\ph_{d} - 2 \ph_{q} = 0 .
\label{23}
\end{align}
Under condition~(\ref{23}), 
mass matrices generally have two more free parameters (for example, $\ph_{q}$ and $\ph_{u} + \ph_{d}$).
Then, the diagonal reflection symmetries can have a similar feature (for $\th_{\rm QFD}$)
to the discrete symmetry $P$ \cite{Mohapatra:1978fy,Beg:1978mt} or $CP$ 
\cite{Nelson:1983zb} in a solution of the strong CP problem. 
Moreover, $\bar \th$ is dynamically retained at zero 
by a flavored axion \cite{Davidson:1981zd, Wilczek:1982rv, Berezhiani:1989fp, Ahn:2014gva, Ema:2016ops,Calibbi:2016hwq} ({\it flaxion} \cite{Ema:2016ops} or {\it axiflavon} \cite{Calibbi:2016hwq}) 
that associates with the SSB of $ U(1)_{\rm PQ}$.
If the cut-off scale $\L$ is taken to be the GUT scale $\L_{\rm GUT} \simeq 10^{16}$ [GeV], 
Eqs.~(\ref{flavonvev1}) and (\ref{flavonvev2}) suggest that 
\begin{align}
\vev{\th_{u,d}} \sim \L_{\rm GUT} {\sqrt{m_{u,d} \, m_{c,s}} \over v} \sim 10^{12} \, [\GeV].  
\end{align}
This is consistent with phenomenological constraints \cite{Ema:2016ops} and 
predicts the axion mass $m_{a} \simeq 10^{-6} \, [\eV]$ and the dark matter abundance $\O_{a} h^{2} \sim 0.2.$ %
%
These chiral and GCP symmetries may shed light 
on the strong CP problem and the origin of the CP violation. 

\section{Physical parameters}

Next, let us consider predictions of mass eigenvalues and mixings. 
Because the four-zero texture can reproduce quark masses and the CKM matrix \cite{Xing:2015sva},  we focus on the lepton sector. 
Derivation of these physical parameters has been performed in a previous study \cite{Yang:2020qsa}. 
In this paper, a precise determination of the Majorana phases is added. 

Diagonalizing the mass matrices $m_{f}^{\rm diag} = U_{Lf}^{\dg} m_{f} U_{Rf}$, 
one obtains an approximate form of the MNS matrix;
\begin{align}
U_{\rm MNS} &= U_{Le}^{\dg} U_{L \n}  \simeq V_{e}^{T} \Diag{-i}{1}{1} V_{\n} ,
\label{UMNS2}
\end{align}
where $V_{\n}$ is an real orthogonal matrix ($V_{\n}^{*} = V_{\n}$) and 
\begin{align}
V_{e} & \simeq 
\begin{pmatrix}
 1 & 0 & 0 \\
 0 & \sqrt{r_{e}} & \sqrt{1-r_{e}} \\[3pt]
 0 & - \sqrt{1-r_{e}} & \sqrt{r_{e}} \\
\end{pmatrix}
\begin{pmatrix}
 1 & \ds - \sqrt{m_{e} \over m_{\m}} & 0 \\
 \ds \sqrt{m_{e} \over m_{\m}} & 1 & 0 \\
 0 & 0 & 1 \\
\end{pmatrix} ,
\label{Vere}
\end{align}
with $r_{e} \equiv A_{e}/m_{\t}$.

The PDG parametrization is written as
\begin{align}
U_{}^{PDG} &= 
\begin{pmatrix}
c_{12} c_{13} & s_{12} c_{13} & s_{13} e^{-i\d_{CP}} \\
-s_{12} c_{23} - c_{12} s_{23} s_{13}  e^{i \d_{CP}} & c_{12} c_{23} - s_{12} s_{23} s_{13} e^{i \d_{CP}} & s_{23} c_{13} \\
s_{12} s_{23} - c_{12} c_{23} s_{13} e^{i \d_{CP}} & -c_{12} s_{23} - s_{12} c_{23} s_{13} e^{ i \d_{CP}} & c_{23} c_{13}
\end{pmatrix} \\
& \times {\rm diag} (1 ,  e^{ i \a_{2} / 2} , e^{ i \a_{3} / 2}) ,
\label{PDG}
\end{align}
where $c_{ij} \equiv \cos \th_{ij}^{PDG}, s_{ij} \equiv \sin \th_{ij}^{PDG}$, 
$\d_{CP}$ is the Dirac phase, and
$ \a_{2}, \a_{3} $ are the Majorana phases. 
The mixing angles and mass differences of the latest global fit \cite{Esteban:2018azc}
\begin{align}
\th_{23}^{PDG} &= 49.7^{\circ} , ~~~~
\th_{12}^{PDG} = 33.82^{\circ} , ~~~~
\th_{13}^{PDG} = 8.61^{\circ} , \\
\D m_{21}^{2} &= 73.9 \, [\meV^{2}], ~~~ 
\D m_{31}^{2} =  2525 \, [\meV^{2}], 
\end{align}
determines the Dirac phase $\d_{CP}$ as
\begin{align}
\sin \d_{CP} &= -0.390 
, ~~~\d_{CP} \simeq 203^{\circ}  .
\end{align}
This is very close to the best fit for the normal hierarchy (NH) $\d_{CP} / ^{\circ} = 217^{+40}_{-28}$ \cite{Esteban:2018azc}.

Next, we proceed to a discussion of the Majorana phases.
The $\m-\t$ reflection symmetry restrict the Majorana phases to be  $\a_{2,3} /2 = n \pi /2$ $(n=0,1)$
\cite{Xing:2017cwb}. 
The nontrivial phase $\pi / 2$ comes from negative mass eigenvalues \cite{Xing:2017cwb, Nath:2018hjx}.
However, the $\m-\t$ reflection symmetries~(\ref{mutausym}) 
no longer retain this property. 
The Majorana phases are located on truly $CP$-violating values. 

The phases are calculated by the rephasing invariants \cite{Branco:1986gr, Jenkins:2007ip, Branco:2011zb}
\begin{align}
I_{1} & = (U_{\rm MNS})_{12}^{2} (U_{\rm MNS})_{11}^{*2} =
{1\over 4}  \sin^{2} 2 \th_{12}^{PDG} \, \cos^{4} \th_{13}^{PDG} \, (\cos \a_{2} + i \sin \a_{2}),  
\label{I1} \\
I_{2} & = (U_{\rm MNS})_{13}^{2} (U_{\rm MNS})_{11}^{*2} =
{1\over 4} \sin^{2} 2 \th_{13}^{PDG} \, \cos^{2} \th_{12}^{PDG}  \, (\cos \a'_{3} + i \sin \a'_{3}),  
\label{I2}
\end{align}
where $\a'_{3} \equiv \a_{3} - 2 \d_{CP}$. 
Substitution of Eq.~(\ref{UMNS2}) into Eqs.~(\ref{I1}) and (\ref{I2}) yields the following results;
\begin{align}
\a_{2}^{0} \simeq 11.3^{\circ}, ~~~ \a_{3}^{0} \simeq 7.54^{\circ}.
\label{alpha0}
\end{align}
As a cross-check, we substituted these results to the PDG parameterization~(\ref{PDG})
and confirmed that the same mixing matrix~(\ref{UMNS2}) were reproduced.

Because Eqs.~(\ref{UMNS2}) and (\ref{alpha0}) do not count contribution from a negative eigenvalue, 
we parameterize these effects as 
\begin{align}
m_{2} = e^{i \b_{2}} |m_{2}| , ~~~ m_{3} = e^{i \b_{3}} |m_{3}| , ~~~ 
\b_{2,3} = 0 ~{\rm or}~ \pi. 
\end{align}
The whole Majorana phases are found to be
\begin{align}
(\a_{2}, \, \a_{3}) = (\a_{2}^{0} + \b_{2}, \, \a_{3}^{0} + \b_{3}) = 
(11.3^{\circ}~{\rm or}~ 191.3^{\circ}, ~ 7.54^{\circ}~{\rm or}~ 187.54^{\circ}). 
\end{align}

Including the Majorana phases, 
one can reconstruct the neutrino mass matrix $m_{\n}$ as 
\begin{align}
m_{\n} = 
V_{e} U_{\rm MNS} \Diag{m_{1}}{m_{2}}{m_{3}} U_{\rm MNS}^{T} V_{e}^{T} .
\label{fullmn}
\end{align}
If the universal texture $(m_{f})_{11} = 0$ for $f = u,d,\n ,e$ \cite{Koide:2002cj} 
and small 2-3 mixing of $V_{e}$ is assumed, we can determine the lightest neutrino mass $m_{1}$
from the condition of the texture
\begin{align}
m_{1} = {-e^{i \alpha_{2}} |m_{2}| s_{12}^{2} - e^{i \alpha_{3}} |m_{3}| t_{13}^{2} \over c_{12}^2  } , 
\end{align}
where $t_{13} \equiv s_{13}/ c_{13}. $ 
The numerical values of the mass are found to be
\begin{align}
|m_{1}| &= 6.20 \, [\meV] ~~  {\rm for} ~~  (\b_{2}, \b_{3}) = (0,0)  \text{ or } (\pi, \pi ) , \\
&= 2.54 \, [\meV] ~~ {\rm for} ~~ (\b_{2}, \b_{3}) = (0, \pi) \text{ or } (\pi, 0)  ,
\end{align}
for the NH case.
For the inverted mass hierarchy, the solutions do not have real values and thus contradict 
the diagonal reflection.

In a previous study \cite{Yang:2020qsa},
the effective mass ${m_{ee}}$ of the double beta decay was also evaluated as
\begin{align}
|m_{ee}| &= \left| \sum_{i=1}^{3} m_{i} U_{ei}^{2} \right| \\
&= 0.17 \, [\meV]  ~~ {\rm for} ~~ (\b_{2}, \b_{3}) = (0,0)  \text{ or } (\pi, \pi ) , \\
 &= 1.24 \, [\meV]  ~~ {\rm for} ~~ (\b_{2}, \b_{3}) = (0, \pi) \text{ or } (\pi, 0)  . 
\end{align}
%

\subsection{Universal four-zero texture} 

Here, we show a universal four-zero texture 
compatible with neutrino mixing parameters. 
An additional assumption in this paper is $(m_{\n})_{13} = 0$.  
This assumption can be justified similar to Eq.~(\ref{13zero}) in the left-right symmetric models. 
This constraint realizes the universal four-zero texture and 
determines the mixing parameter $r_{e} = A_{e}/m_{\t}$ in Eq.~(\ref{Vere}).
%

The mass matrix $m_{\n}$~(\ref{fullmn}) is a matrix function of $\a_{2}, \a_{3}, m_{1},$ and $r_{e}$. 
Solving an equation $(m_{\n})_{13} =0$, we find two solutions for universal four-zero texture.
The first solution with a large $r_{e} \simeq 0.996$ and its mass eigenvalues are found to be
\begin{align}
m_{\n 0} \simeq &
\begin{pmatrix}
0 & -8.86 i &  0 \\
- 8.86 i & 29.3 & 26.4 \\
0 & 26.4 & 14.6
\end{pmatrix} [\meV]  
~~~
 {\rm for} ~~ (\a_{2}, \a_{3}) = (\pi, 0),  \label{mn0}  \\
&(m_{1} \, , m_{2} \, ,  m_{3} ) = (2.54 ,  \,  -8.96 ,  \, 50.3) \, [\meV] . 
\end{align}
Indeed, the Majorana phases $\b_{2} = \pi , \b_{3} = 0$ are realized. 
In this basis, the charged lepton mass matrix also shows the four-zero texture
\begin{align}
m_{e} &\simeq  
\begin{pmatrix}
 0 & -7.058 & 0 \\
 -7.058 & 107.873 & 96.12 \\
 0 & 96.12 & 1740 \\
\end{pmatrix} \, [\MeV] ~~~  {\rm for} ~~  (m_{e}^{\rm diag} )_{11} < 0 , \, (m_{e}^{\rm diag} )_{22} > 0 \, , \\
&\simeq
\begin{pmatrix}
 0. & 7.058 & 0 \\
 7.058 & -95.898 & 108.1 \\
 0 & 108.1 & 1740 \\
\end{pmatrix} \, [\MeV] ~~~  {\rm for} ~~ (m_{e}^{\rm diag} )_{11} > 0  , \, (m_{e}^{\rm diag} )_{22} < 0) \, .
\end{align}
The second solution has a small $r_{e} \simeq 0.0024$;
\begin{align}
\tilde m_{\n 0} = &
\begin{pmatrix}
0 & 10.5 \, i & 0 \\
10.5 \,  i & 24.9 & -22.0 \\
0 & -22.0 & 30.1 \\
\end{pmatrix} \, [\meV] 
~~~
 {\rm for} ~~ (\a_{2}, \a_{3}) = (0, 0),  \\
 &(m_{1} \, , m_{2} \, ,  m_{3} ) = (- 6.20 ,  \,  10.6 ,  \, 50.6) \, [\meV] . 
\end{align}
%
 This solution results in  $(m_{e})_{22} \simeq m_{\t}$ and
seems to be somewhat unnatural.  
However, it may relate large 22 and 23 elements of quarks 
Eq.~(\ref{bestfit1}) and~(\ref{bestfit2}) by a grand unified theory (GUT).


The right-handed neutrino mass matrix $M_{R}$ can be 
reconstructed from the type-I seesaw mechanism \cite{Minkowski:1977sc, Yanagida:1979as, Mohapatra:1979ia, GellMann:1980v} with some GUT relations. 
A $u-\n$ unification, such as in the Pati--Salam GUT \cite{Pati:1974yy}, can determine $Y_{\n}$ 
from Eq.~(\ref{bestfit1}) as
\begin{align}
Y_{\n} = 
Y_{u} \simeq {0.9 m_{t} \sqrt 2 \over v}
\begin{pmatrix}
0 & 0.0002 \, i & 0 \\
- 0.0002 \, i & 0.10 & 0.31 \\
0 & 0.31 & 1\\
\end{pmatrix} .
\label{53}
\end{align}

From Eq.~(\ref{mn0}) and~(\ref{53}), 
$M_{R}$ also displays a four-zero texture 
because the four-zero texture is seesaw invariant \cite{Nishiura:1999yt,Fritzsch:1999ee}, 
\begin{align}
& M_{R} = {v^{2} \over 2} Y_{\n} m_{\n 0 }^{-1} Y_{\n}^{T} \\ 
& = 
\begin{pmatrix}
0 & -1.08 \, i \times 10^{8} & 0 \\
-1.08 \, i \times 10^{8} & 1.26 \times 10^{14} & 4.07 \times 10^{14} \\
0 & 4.07 \times 10^{14} & 1.32 \times 10^{15}
\end{pmatrix} [\GeV] .
\label{MR}
\end{align}
Evidently, $M_{R}$ also satisfies diagonal reflection symmetry~(\ref{refsym}),
\begin{align}
R M_{R}^{*} R = M_{R}. 
\end{align}
Therefore,  all the fermion masses respect the diagonal reflection symmetry 
with a four-zero texture.

The eigenvalues of $M_{R}$ are found to be
\begin{align}
& (M_{R1} \, , M_{R2} \, , M_{R3}) \nn \\ &= (2.86 \times 10^{6} \, , 3.73 \times 10^{9} \, , 1.44 \times 10^{15}) \, [\GeV] . \label{41} 
\end{align}
The Yukawa matrix $Y_{\n}$~(\ref{53}) is evaluated at $m_{Z}$ scale. 
Other renormalized values of quark masses will lead to smaller eigenvalues of $ M_{R}$.
For example, $Y_{\n}$ is determined in other Pati--Salam GUT 
\begin{align}
Y_{\n} = 
\Diag{i}{1}{1}
\begin{pmatrix}
0 & C_{\n} & 0 \\ 
 C_{\n} & \tilde B_{\n} & B_{\n} \\
0 & B_{\n}  & A_{\n} 
\end{pmatrix}
\Diag{-i}{1}{1} , 
\end{align}
with $A_{\n} = A_{u}, C_{\n} = C_{u}$ and the Georgi--Jarlskog relation $B_{\n} = - 3 B_{u}, \tilde B_{\n} = - 3 \tilde B_{u}$ \cite{Georgi:1979df}.
Quark masses at the GUT scale $\L_{\rm GUT} = 2 \times 10^{16} \, $[GeV] \cite{Xing:2007fb}
\begin{align}
m_u = 0.48 \, [\MeV] , ~
m_c = 0.235 \, [\GeV], ~ 
m_t = 74 \, [\GeV] ,
\end{align}
lead to smaller eigenvalues 
\begin{align}
& (M_{R1} \, , M_{R2} \, , M_{R3}) \nn \\ &= (9.18 \times 10^{4} \, , 1.77 \times 10^{9} \, , 3.02 \times 10^{14}) \, [\GeV] \label{44} . 
\end{align}
The precise eigenvalues will be obtained by solving renormalization group equations.

The mass matrix $M_{R}$ is constrained by 
the diagonal reflection symmetries, the universal four-zero texture, and the type-I seesaw mechanism. This scheme enhances the predictivity of leptogenesis \cite{Fukugita:1986hr}. 
Large $CP$ violation in $M_{R}$ (and $m_{\n}$) is desirable.

Because the mass matrix $M_{R}$ has strong hierarchy $M_{R} \sim Y_{u}^{T} Y_{u}$, 
the lightest mass eigenvalue $M_{R1}$ is too small \cite{Davidson:2002qv, Hamaguchi:2001gw}  
for naive thermal leptogenesis. 
However, leptogenesis may be achieved by the decay of the second lightest neutrino $\n_{R2}$ \cite{Vives:2005ra}
with the maximal Majorana phase $\a_{2} /2 \sim \pi/2$.

\section{Quantum corrections} 

Here we show the stability of the symmetries against quantum corrections. 
Because quantum corrections are very small for the first generation, 
the symmetries~(\ref{refsym}) are retained as approximate ones.

The diagonal reflection symmetries are not invariant under the renormalization group equations (RGEs) of the SM. 
RGEs of quarks at one-loop order are given by \cite{Xing:2019vks},
\begin{align}
16 \pi^{2} {d Y_{u} \over dt} &= [\a_{u} + C_{u}^{u} (Y_{u} Y_{u}^{\dg}) + C_{u}^{d} (Y_{d} Y_{d}^{\dg}) ] Y_{u} , \label{RGEu} \\
16 \pi^{2} {d Y_{d} \over dt} &= [\a_{d} + C_{d}^{u} (Y_{u} Y_{u}^{\dg}) + C_{d}^{d} (Y_{d} Y_{d}^{\dg}) ] Y_{d} , \label{RGEd}
\end{align}
where $t = \ln (\m) / m_{Z}$, $\m$ is an arbitrary renormalization scale, 
$\a_{f}$ are flavor independent contributions from the gauge and Higgs bosons. 
The coefficients $C_{f}^{f'}$ are given by 
\begin{align}
C_{u}^{d} = C_{d}^{u} = - 3/2 , ~~ C_{u}^{u} + C_{d}^{d} = 3/2 . 
\end{align}
Similar equations hold in the lepton sector.

It has been pointed out that the four-zero texture and its CKM phase are approximately RGE invariant \cite{Fritzsch:1997fw, Xing:2015sva}. The same statement holds for the diagonal reflection.
Some of the best fit values~(\ref{bestfit1}) and (\ref{bestfit2}) can be roughly written as
\begin{align}
Y_{u} &
\simeq {\sqrt 2 \over v}
\begin{pmatrix}
0 & i \sqrt{m_{u} m_{c}} & 0 \\
-i \sqrt{m_{u} m_{c}} & O(m_{t}) & O(m_{t}) \\
0 & O(m_{t}) & O(m_{t}) \\
\end{pmatrix} , \\
Y_{d} &
\simeq {\sqrt 2 \over v}
\begin{pmatrix}
0 & \sqrt{m_{d} m_{s}} & 0 \\
\sqrt{m_{d} m_{s}} & O(m_{b}) & O(m_{b}) \\
0 & O(m_{b}) & O(m_{b}) \\
\end{pmatrix} .
\end{align}
A term in Eq.~(\ref{RGEd}) can be reconstructed as
\begin{align}
Y_{u} Y_{u}^{\dg} Y_{d} & = 
\begin{pmatrix}
1.17 \times 10^{-9} i & 2.34 \times 10^{-12} + 2.56 \times 10^{-7} i & 7.99 \times 10^{-7} i \\
 6.22 \times 10^{-6} & 0.00140 - 1.17 \times 10^{-9} i & 0.00438 \\
 2.00 \times 10^{-5} & 0.00450 - 3.63 \times 10^{-9} i & 0.0141 \\
\end{pmatrix} 
\\ & \simeq
\begin{pmatrix}
i C_{u} \tilde B_{u} C_{d}  & i C_{u} (B_{u} B_{d}  + \tilde B_{u} \tilde B_{d}) & 
 i C_{u} (B_{u} A_{d}  + \tilde B_{u} B_{d}) \\ 
( B_{u} B_{u} + \tilde B_{u} \tilde B_{u} ) C_{d}  &  O( B_{u} A_{u} B_{d}) -i \tilde B_{u} C_{u} C_{d} & O(B_{u} A_{u} A_{d} )  \\ 
(A_{u} B_{u} + B_{u} \tilde B_{u}) C_{d} & O( A_{u} A_{u} B_{d})  -i B_{u} C_{u} C_{d} & O (A_{u} A_{u} A_{d})
\end{pmatrix} . \label{55}
\end{align}
In Eq.~(\ref{55}), several terms at the leading order are represented. 
Matrix elements of the first row and column (specifically, $(1, i)$ and $(j, 1)$ elements) of the term $Y_{u} Y_{u}^{\dg} Y_{d}$ are insignificant.
This is due to the smallness of $|(m_{u,d})_{12}| = |C_{u,d}| \simeq \sqrt{m_{u,d} m_{c,s}}$ 
(or the chiral symmetry of the first generation $U(1)_{\rm PQ}$). 
Furthermore, the influence of complex phases of $(2, 2), (2,3), (3,2)$ and  $(3, 3)$ elements are also negligible because they are the second-order corrections  of the small parameters $C_{u,d}$. 

Because the flavor dependent terms in Eqs.~(\ref{RGEu}) and (\ref{RGEd}) 
have a similar structure, flavor dependent contributions hardly change 
the couplings of the first generation. 
This statement holds without the four-zero texture 
as long as couplings in the first row and column of the Yukawa matrices are sufficiently small.
Therefore, the diagonal reflection symmetries with these properties 
are approximately RGE invariant
and inherit flavor structures at a high energy scale.

\section{Summary} 

In this paper, we considered a set of new symmetries in the SM: {\it diagonal reflection} symmetries.
$\m-\t$ reflection symmetries from a previous study are deformed to 
$R \, m_{u,\nu}^{*} \, R = m_{u,\nu}, ~ m_{d,e}^{*} = m_{d,e}$ with $R =$ diag $(-1,1,1)$ 
by a redefinition of fermion fields. 
They can constrain the Majorana phases to be $\a_{2,3} /2 \sim 0$ or $\pi /2$ and enhance the predictivity of  leptogenesis. 

The form of the symmetries suggests that the flavored $CP$ violation only comes from 
a chiral symmetry breaking of the first generation. 
As a justification of  
diagonal reflection symmetries and a zero texture $(m_{f})_{11} = 0$,  
simultaneous breaking of a chiral $U(1)_{\rm PQ}$ and a generalized $CP$ symmetry is discussed 
in a specific 2HDM.
As a result, 
a flavored axion appears in conjunction with solving the strong CP problem.
The axion scale is suggested to be $\vev{\th_{u,d}} \sim \L_{\rm GUT} \, \sqrt{m_{u,d} \, m_{c,s} } / v \sim 
10^{12} \, $[GeV]. 
This value can produce  the dark matter abundance $\O_{a} h^{2} \sim 0.2$ and is very intriguing. 
They can be also applicable to a solution of the strong CP problem by discrete symmetry of $P$  or $CP$  
because the symmetries can reconcile the CKM phase $\d_{\rm CKM}$ and $\th_{\rm QFD}^{\rm tree} = {\rm Arg} \, \Det[m_{u} m_{d}] = 0$ without Hermiticity or mirror fermions.

By combining the symmetries with the four-zero texture, 
the mass eigenvalues and mixing matrices of quarks and leptons are reproduced well. 
This scheme predicts the normal hierarchy, the Dirac phase $\d_{CP} \simeq 203^{\circ},$  and $|m_{1}|\simeq 2.5$ or $6.2 \, [\meV]$.

The type-I seesaw mechanism results in the mass matrix of the right-handed neutrinos $M_{R}$, which exhibits diagonal reflection symmetries with a four-zero texture. 
The matrix $M_{R}$ is completely determined by 
a given $Y_{\n}$ and the type-I seesaw mechanism. 
%
A $u-\n$ unification predicts that the mass matrix $M_{R}$ has a strong hierarchy $M_{R} \sim Y_{u}^{T} Y_{u}$. 

The symmetries are approximately stable under the renormalization of SM. 
This statement holds without the four-zero texture as long as couplings in the first row and column of the Yukawa matrices are sufficiently small.
Then, they can possess information on a high energy scale. 


\section*{Acknowledgment}

This study is financially supported 
by JSPS Grants-in-Aid for Scientific Research
No.~JP18H01210, No. 20K14459,  
and MEXT KAKENHI Grant No.~JP18H05543.


\end{document}